\documentclass[10pt,a4paper]{article}
\usepackage{amsmath}
\usepackage{multirow}
\usepackage{amssymb}
\usepackage{amsfonts}
\usepackage{amstext}
\usepackage{amsbsy}
\usepackage[mathscr]{eucal}
\usepackage{graphicx}
\usepackage{color}
\usepackage{epstopdf}
\usepackage{verbatim}
\usepackage[all]{xy}
\newcommand{\beq}{\begin{equation}}
\newcommand{\eeq}{\end{equation}}
\newcommand{\beqa}{\begin{eqnarray}}
\newcommand{\eeqa}{\end{eqnarray}}

 \numberwithin{equation}{subsection}

\makeindex
\title{\Large\textbf{Quantum solution to a three player Kolkata restaurant problem using entangled qutrits}}

\author{\textit{Puya Sharif$^{ \dagger}$ and Hoshang Heydari }\\
        \small\textit{Department of physics, Stockholm university 10691 Stockholm Sweden}\\
\\\small\textit{$\dagger$ Email: ps@puyasharif.net}}
\begin{document}
\maketitle

\begin{abstract}
Three player quantum Kolkata restaurant problem is modeled using three entangled qutrits. This first use of three level quantum states in this context is a step towards a $N$-choice generalization of the $N$-player quantum minority game.  It is shown that a better than classical payoff is achieved by a Nash equilibrium solution where the space of available strategies is spanned by subsets of SU(3) and the players share a tripartite entangled initial state. 
\end{abstract}

\section{Introduction}
Quantum game theory is a fairly recent extension of game theoretical analysis to situations formulated in the framework of quantum information theory. The first papers appeared in 1999. Meyer showed with a model of a penny-flip game that a player making a \emph{quantum move} always comes out as a winner against a player making a \emph{classical} move regardless of the classical players choice \cite{meyer}. The same year Eisert et al. published a quantum protocol in which they overcame the dilemma in Prisoners dilemma \cite{eisert}. In 2003 Benjamin and Hayden generalized Eisert's protocol to handle multiplayer quantum games and introduced the quantum minority game together with a solution for the four player case which outperformed the classical randomization strategy \cite{benjamin}. This result was later generalized to the $n$-players by Chen et al. in 2004 \cite{chen}. Multiplayer minority games has since then been extensively investigated by Flitney et al. \cite{flitney1,flitney2,schmid}.

We will here extend quantum minority games to situations where there are not only multiple players, but also multiple choices. A quantum version of the Kolkata restaurant problem, which is a generalized minority game will be presented. The players uses maximally entangled qutrits as a quantum resource and selects their strategy by locally acting with a general SU(3) operator on the qutrit in their possession.
\subsection{Kolkata restaurant problem}
The Kolkata restaurant problem is a minority-type game \cite{kolkata,kolkata1,kolkata2,kolkata3,Elfarol}. In its most general form $N$ non-communicating agents (players), have to choose between  $n$ choices. The agents receive a gain in their utility if their choice is not too crowded, i.e the number of agents that made the same choice is under some threshold limit. The choices can also have different values of utility associated with them, accounting for a preference profile over the set of choices. The original formulation comes with a story of  workers in Kolkata that during lunch hours has to choose between a fixed number of cheap restaurants. Each restaurant can only serve a finite number of customers, so workers arriving to a crowded restaurant will simply miss the opportunity of having lunch. Often is the number of agents taken to be equal to the number of restaurants, and the maximum number of costumers per restaurant limited to one. The problem is usually modeled as an iterative game where agents ought to base their decision on information about the distribution of agents over choices in the previous iterations. The Kolkata restaurant problem offers therefore a method for modeling heard behavior and market dynamics, where visiting a restaurant translates to buying a security, in which case an agent wishes to be the only bidder.
\subsection{The model}
In our simplified model there are just three agents, Alice, Bob and Charlie. They have three possible choices: security 0, security 1 and security 2. They receive a payoff $\$$ of one unit if their choice is unique, i.e that nobody else has made the same choice, otherwise they receive $\$=0$. The game is so called \emph{one shoot}, which means that it is non-iterative, and the agents have no information from previous rounds to base their decisions on. Under the constraint that they cannot communicate, there is nothing left to do other than randomizing between the choices. Given the symmetric nature of the problem, any deterministic strategy would lead all three agents to the same strategy, which in turn would mean that all three would leave empty handed. There are $27$ different strategy profiles possible, i.e combinations of choices. $12$ of which gives a payoff of $\$=1$ to each one of them. Randomization gives therefore agent $i$ an expected payoff of $E^c(\$)=\frac{4}{9}$, where the superscript denotes that the result is due to the best \emph{classical} strategy (as opposed to \emph{quantum} strategy).

In the framework of quantum game theory \cite{flitney,pitrowski1,NEinQ,landsburg}, Alice, Bob and Charlie shares a quantum resource. Each has a part of a multipartite quantum state. They play their strategy by manipulating their own part of the combined system, before measuring their subsystems and choosing accordingly. Whereas classically the players would be allowed randomizing over a discrete set of choices,  in the quantum version each subsystem is allowed to be transformed with the full machinery of quantum operations. A strategy, or choice therefore translates to choosing a unitary operator $U$. In the absence of entanglement, quantum games of this type usually yield the same payoffs as their classical counterparts, whereas the combination of unitary operators (or a subset therein) and entanglement, sometimes strongly outperforms classical games and decision theoretic models. We will here present such a case.

\section{Qutrits and parametrization of SU(3)}

A qutrit is a 3-level quantum system on 3-dimensional Hilbert space $\mathcal{H}=\mathbb{C}^{3}$ , written in the computational basis as:

\begin{equation}
  |\psi\rangle=a_{0}|0\rangle+a_{1}|1\rangle+a_{2}|2\rangle \in \mathbb{C}^3,
\end{equation}
with $a_{0},a_{1},a_{2}\in\mathbb{C}$  and $|a_{0}|^{2}+|a_{1}|^{2}+|a_{2}|^{2}=1 $. A general $N$-qutrit system $\left|\Psi\right\rangle$   is a vector on $3^{N}$-dimensional Hilbert space, and is written as a linear combination of $3^{N}$ orthonormal basis vectors.

\begin{equation}
  \left|\Psi\right\rangle =\sum_{x_{N},..,x_{1}=0}^{2}a_{x_{N}...x_{1}}\left|x_{N}\cdots x_{1}\right\rangle ,
\end{equation}
where
\begin{equation}
\left|x_{N}\cdots x_{1}\right\rangle =\left|x_{N}\right\rangle \otimes\left|x_{N-1}\right\rangle \otimes\cdots\otimes\left|x_{1}\right\rangle \in\mathcal{H}=\overbrace{\mathbb{C}^{3}\otimes...\otimes\mathbb{C}^{3}}^\text{$N$-times},
\end{equation}
with $x_{i}\in\{0,1,2\}$  and complex coefficients $a_{x_{i}}$,  obeying $\sum|a_{x_{N}...x_{1}}|^{2}=1$.

Single qutrits are transformed with unitary operators $U \in$ SU(3), i.e operators from the special unitary group of dimension 3, acting on $\mathcal{H}$ as $U:\mathcal{H} \rightarrow \mathcal{H}$. In a multi-qutrit system, operations on single qutrits are said to be local. They affect the state-space of the corresponding qutrit only. The transformation of a multi-qutrit state vector under local operations is given by the tensor products of the individual operators:

\begin{equation}
\left|\Psi_{fin}\right\rangle =U_{N}\otimes U_{N-1}\otimes\cdots\otimes U_{1}\left|\Psi_{in}\right\rangle,
\end{equation} where $\left|\Psi_{in}\right\rangle$ and $\left|\Psi_{fin}\right\rangle$ denotes the initial and final state of the system respectively.

There are a number of ways you can parameterize SU(3) \cite{on(su3),quotient}. One common approach is through the Lie algebra of the group, the eight traceless $3\times3$  Gell-Mann matrices. We are using a different and maybe slightly more intuitive parametrization \cite{coherent}. Let $\bar{x}, \bar{y}, \bar{z}$  be three general, mutually orthogonal complex unit vectors, such that $\bar{x}\cdot\bar{y}=0$  and $\bar{x}^*\times\bar{y}=\bar{z}$. We construct a SU(3) matrix by placing $\bar{x}, \bar{y}^*$ and $\bar{z}$  as its columns.
Now a general complex unit vector is given by:

\begin{equation}
\bar{x}=\left(\begin{array}{c}
\sin\theta\cos\phi e^{i\alpha_{1}}\\
\sin\theta\sin\phi e^{i\alpha_{2}}\\
\cos\theta e^{i\alpha_{3}}
\end{array}\right),
\end{equation}
and one complex unit vector orthogonal to $\bar{x}$  is given by:

\begin{equation}
\bar{y}=\left(\begin{array}{c}
\cos\chi\cos\theta\cos\phi e^{i(\beta_{1}-\alpha_{1})}+\sin\chi\sin\phi e^{i(\beta_{2}-\alpha_{1})}\\
\cos\chi\cos\theta\sin\phi e^{i(\beta_{1}-\alpha_{2})}-\sin\chi\cos\phi e^{i(\beta_{2}-\alpha_{2})}\\
-\cos\chi\sin\theta e^{i(\beta_{1}-\alpha_{3})}
\end{array}\right),
\end{equation}
where $0\leq\phi,\theta,\chi,\leq\pi/2$ and $0\leq\alpha_1,\alpha_2,\alpha_3,\beta_1,\beta_2\leq2\pi$. We have a general SU(3) matrix $U$, given by:

\begin{equation}
U=\left(\begin{array}{ccc}
x_{1} & y_{1}^{*} & x_{2}^{*}y_{3}-y_{3}^{*}x_{2}\\
x_{2} & y_{2}^{*} & x_{3}^{*}y_{1}-y_{1}^{*}x_{3}\\
x_{3} & y_{3}^{*} & x_{1}^{*}y_{2}-y_{2}^{*}x_{1}
\end{array}\right),
\end{equation}
and it is controlled by eight real parameters ${\phi,\theta,\chi,\alpha_1,\alpha_2,\alpha_3,\beta_1,\beta_2}$. 

\section{The scheme}
The scheme under study is a development of one first introduced by Eisert et al. \cite{eisert}, and later generalized by Benjamin and Hayden \cite{benjamin}. It starts out with Alice, Bob and Charlie, $A,B$ and $C$  respectively, sharing a quantum resource, an entangled tripartite 3-level quantum state. We need to allow the quantum states to have a common origin, since creating entanglement is a global operation, and can´t be done by acting locally on the subsystems. We assume that there exists an unbiased referee that prepares the state and distributes the subsystems among the players. From that point on, no communication is allowed between the players and the referee.
 Each qutrit is due to be measured by the player owning it, at the end of the protocol in the $\{|0\rangle,|1\rangle,|2\rangle\}$ -basis, where basis vector corresponds to one of the three choices: security 0, security 1, and security 2. The players plays their strategy by applying an operator from the set of allowed strategies $S$, followed by a local measurement which determines their final choice. The unitary operations done by $A,B,C$ are done locally, which means that the operator is applied on the subsystem held by the player. As mentioned, this translates to the transformation of $|\psi_{in}\rangle$ by the tensor product of the unitary operators applied by the players.

 We want to create a quantization of the classical game in which we expand the set of available strategies to include quantum moves. While we are proposing a quantum game which in some sense is fundamentally different from the classical version, we require it to be an extension, not an addition to the classical Kolkata restaurant problem. Tracing the steps of the predecessors of this protocol, we restrict our formulation to have the classical game fully present at all times, accessible in the form of restrictions on the set of allowed local operations. We simply require that there exists a set of operators that when applied locally on an entangled initial state gives the same outcomes as in the classical non-quantum version.
 Lets first look at the classical game presented with quantum formalism. Note that there is nothing quantum mechanical happening at this point. The initial state $|\psi_{in}\rangle=|000\rangle=|0\rangle_{C}\otimes|0\rangle_{B}\otimes|0\rangle_{A}$ corresponds to the case where the three players chooses security 0, by default. The individual choices are made by applying operators $s_i,s_j,s_k \in S=\{s_0,s_1,s_2\}$ to each subsystem. The exact form of these operators can be left to discuss later. The only restriction at the moment is that they obey: $s_0|0\rangle = |0\rangle,\,s_1|0\rangle = |1\rangle,\,s_2|0\rangle = |2\rangle$, resulting in fully deterministic outcomes:
 \begin{equation}
 s_i\otimes s_j \otimes s_k|000\rangle=|i\, j\, k\rangle.
 \end{equation}
 As mentioned earlier there are 27 different such outcomes, each linked to different combinations of the operators $\{s_0,s_1,s_2\}$, 12 of which gives a player a payoff $\$=1$, and the rest $\$=0$. Clearly there are no operators available corresponding to mixed strategies, so randomization processes leading to classical mixed strategies are here lifted outside the protocol and is done by the players before the choices are made.  Having finished the first step in the quantization process our task is now to keep the classical game present throughout the coming steps while we add quantum structure by choosing an entangled initial state and expanding the set of strategy operators to include any $U \in$ $S$ = SU(3).
Since the game is symmetric and unbiased in regards to permutation of player positions, then this is a property that has to be true of the initial state $|\psi_{in}\rangle$, to assure that the payoff functions $\$_i(|\psi_{in}\rangle,U_A,U_B,U_C)$ of all three players $i$ are identical up to some permutation of $U_A,U_B,U_C$. Note that when dealing with mixed classical and quantum strategies the payoff function becomes an expectation value $E(\$)$ of a probability distribution over the different outcomes. We summarize the criteria for choosing an initial state:
\begin{enumerate}
  \item  The state ought to be entangled, to accommodate for correlated randomization among the players.
  \item  It should be symmetric and unbiased in regards to player positions.
  \item  It must allow for classical game to be accessed by restrictions on the space of available strategy operators.
\end{enumerate}

The three qutrit GHZ-type-state:
\begin{equation}\label{eq:GHZ}
\mid\psi_{in}\rangle=\frac{1}{\sqrt{3}}\left(|000\rangle+|111\rangle+|222\rangle\right),
\end{equation}
not only fulfills the above criteria, it is also a \emph{maximally} entangled state on $\mathcal{H}=\mathbb{C}^{3}\otimes\mathbb{C}^{3}\otimes\mathbb{C}^{3}$. It has the additional property of initializing the game in an maximally undesired state. i.e. one in which none of the players receives any payoff. In order to change their situation, they will have to make an active choice.
It is left to show that we can define a set of operators corresponding to classical pure strategies that gives raise to deterministic classical payoffs when applied to the entangled initial state. This problem was addressed by Eisert et. al. \cite{eisert}, and further developed by Benjamin et.al. \cite{benjamin} for cases of $n$ players and two choices, by defining an entangling operator $J$ and its inverse $J^{\dagger}$, acting on a $n$-\emph{qubit} product state $|00\cdots 0\rangle$ with Hermitian strategy operators $\hat{s_0},\hat{s_1}$, sandwiched in between. By showing that any combination of the classical strategies $\hat{s}_{x_1}\otimes \hat{s}_{x_2}\otimes \cdots \otimes \hat{s}_{x_n}, x_i \in \{0,1\}$ commutes with $J$, one guarantees that the classical game is embedded in the quantum version. That route is not possible when formulating a game with aid of higher dimensional quantum states like qutrits, since at least \emph{two different} Lie-algebra elements of su(3) must appear in the Hamiltonian of $J$ (For the GHZ-type-state), whereby commutation is no longer a fact in the general case. We need a set of operators that replicates the classical payoffs when applied directly on our entangled initial state $\mid\psi_{in}\rangle$.

The cyclic group of order three, $C_3$, generated by the matrix:

 \begin{equation}
 s=\ \left( \begin{array}{ccc}
0 & 0 & 1 \\
1 & 0 & 0 \\
0 & 1 & 0 \end{array} \right)\ ,
 \end{equation}
where $s^3=s^0=I$ and $s^2=s^T$, has the properties we are after. The set of classical strategies $S=\{s^0,s^1,s^2\}$ with $ s^i\otimes s^j \otimes s^k|000\rangle=|i\, j\, k\rangle$ acts on the GHZ-state as:

\begin{multline}
s^i\otimes s^j \otimes s^k\frac{1}{\sqrt{3}}\left(|000\rangle+|111\rangle+|222\rangle\right)=\\
= \frac{1}{\sqrt{3}}\left(|0+i\;0+j\;0+k\rangle+|1+i\;1+j\;1+k\rangle+|2+i\;2+j\;2+k\rangle\right).
\end{multline}
Note that the superscripts denotes powers of the generator and that the addition is modulo 3. In the case under study, where there is no preference profile over the different choices, any combination of the operators in $S=\{s^0,s^1,s^2\}$ leads to the same payoffs when applied to the GHZ-state as to $|000\rangle$.

Now that an entangled initial state $\mid\psi_{in}\rangle$ is chosen, the scheme for the quantum game proceeds as follows.
We form a density matrix $\rho_{in}$  out of the initial state $\mid\psi_{in}\rangle$ and add noise that can be controlled by the parameter $f$ \cite{schmid}. We get:
\begin{equation}
\rho_{in}=f\mid\psi_{in}\rangle\langle\psi_{in}\mid+\frac{1-f}{27}\mathbb{I_{\mathrm{27}}},
\end{equation}
 where $\mathbb{I_{\mathrm{27}}}$ is the $27\times 27$ identity matrix. Alice, Bob and Charlie now applies a unitary operator $U$ that maximizes their chances of receiving a payoff $\$ =1$, and thereby the initial state $\rho_{in}$  is transformed into the final state $\rho_{fin}$.
\begin{equation}
\rho_{fin}=U^{\dagger}\otimes U^{\dagger}\otimes U^{\dagger}\rho_{in}U\otimes U\otimes U.
 \end{equation}
 Note that they are all applying the same operator $U$ since in the absence of communication, coordination of which operator to be applied by whom, would be impossible. We define for each player $i$  a payoff-operator $P_{i}$ , which contains the sum of orthogonal projectors associated with the states for which player $i$  receives a payoff $\$=1$ . For Alice this would correspond to
 \begin{multline}
 P_{A}=\left(\sum_{x_3,x_2,x_1=0}^{2}|x_{3}x_{2}x_{1}\rangle\langle x_{3}x_{2}x_{1}|,\, x_3\neq x_2, x_3\neq x_1, x_2\neq x_1\right)+\\
 +\left(\sum_{x_3,x_2,x_1=0}^{2}|x_{3}x_{2}x_{1}\rangle\langle x_{3}x_{2}x_{1}|,\, x_3=x_2\neq x_1\right).
 \end{multline}
 The expected payoff $E_{i}(\$)$  of player $i$  is calculated by taking the trace of the product of the final state $\rho_{fin}$  and the payoff-operator $P_{i}$:
\begin{equation}
E_{i}(\$)=\mathrm{Tr\left(\mathit{P}_{i}\rho_{fin}\right)}.
\end{equation}

\section{Optimal strategy}
The problem now is to find the unitary operator $U(\phi,\theta,\chi,\alpha_1,\alpha_2,\alpha_3,\beta_1,\beta_2)$ that maximizes the expected payoff. Due to the symmetry of the problem, optimization can be done with respect to the $P_{i}$ of any of the three players. Doing so one arrives at a maximum expected payoff of $E(\$)=\frac{6}{9}$, assuming ($f=1$), compared to the classical $E^c(\$)=\frac{4}{9}$. Which is an $50\%$ increase. This occurs when the players applies the optimal unitary operator  $U^{opt}$, whose parameters are listed in table 1.

\begin{table}\center
\label{Uopttable}
\begin{tabular}{|c|c|c|c|c|c|c|c|c|c|}
\hline
\hline
 Parameter  & $\phi$ & $\theta$  & $\chi$ & $\alpha_1$ & $\alpha_2$ & $\alpha_3$ & $\beta_1$ & $\beta_2$\tabularnewline

\hline
 Value  & $\frac{\pi }{4}$ & $\cos ^{-1}\left(\frac{1}{\sqrt{3}}\right)$ & $\frac{\pi }{4}$ & $\frac{5 \pi }{18}$ & $\frac{5 \pi }{18}$ & $\frac{5 \pi }{18}$ & $\frac{\pi }{3}$ & $\frac{11 \pi }{6}$ \tabularnewline
\hline
\end{tabular}
\caption{$U^{opt}$ in the given parametrization.}
\end{table}

Because of the periodic nature of the solution, there could be more than one unique choice for some of the parameters within the allowed ranges $0\leq\phi,\theta,\chi\leq\pi/2$ and $0\leq,\alpha_1,\alpha_2,\alpha_3,\beta_1,\beta_2\leq2\pi$. This is the case for $\alpha_1,\alpha_2,\alpha_3$, where maximum expected payoff is achieved for $ \left(\frac{5+12n}{18}\right)\pi$, $n\in \{0,1,2\}$. Noting that the the center of SU(3), Z(3) =$\{I,e^{\pm\frac{i 2\pi}{3}}I\}$ only adds a global phase and leaves the density matrix invariant, one concludes that the transformation belongs to SU(3)/Z(3). This removes the above ambiguity, ending up with $\alpha_1,\alpha_2,\alpha_3 = \frac{5 \pi }{18}$.

The final state arrived at by playing $U^{opt}$ is given by:
\begin{multline}
\mid\psi_{fin}\rangle=\frac{1}{3}\left(|000\rangle+|012\rangle+|021\rangle+|102\rangle\right.+\\\left.|111\rangle+|120\rangle+|201\rangle+|210\rangle+|222\rangle\right).
\end{multline}
This is an even distribution of all the states that leads to payoff to all three players and the states which gives payoff to none and shows that the $U^{opt}\otimes U^{opt}\otimes U^{opt}$-operation fails to make the state  fully depart from the space spanned by $|000\rangle,|111\rangle,|222\rangle$. This failure accounts for the expected payoff not reaching unity.

Now by setting $\alpha_1,\alpha_2=0$ and $\alpha_3=\alpha$, in the parametrization, one arrives at a six parameter subset of SU(3), given by operators $U(\phi,\theta,\chi,\alpha,\beta_1,\beta_2)$. The optimum is at the same value as with the transformation belonging to its domain. There is thereby a $V^{opt}$ in this subset, given in table 2 below, that gives each player an expected payoff of $E(\$)=\frac{6}{9}=\frac{2}{3}$.

\begin{table}[!h]\center
\label{Uopttable}
\begin{tabular}{|c|c|c|c|c|c|c|c|c|c|}
\hline
\hline
 Parameter  & $\phi$ & $\theta$  & $\chi$ & $\alpha$ & $\beta_1$ & $\beta_2$\tabularnewline

\hline
 Value  & $\frac{\pi }{4}$ & $\cos ^{-1}\left(\frac{1}{\sqrt{3}}\right)$ & $\frac{\pi }{4}$ & $\frac{\pi }{2}$ & $\frac{\pi }{3}$ & $\frac{5 \pi }{6}$ \tabularnewline
\hline
\end{tabular}
\caption{$V^{opt}$ in the reduced parametrization.}
\end{table}

$U^{opt}$ and $V^{opt}$ differs only by a constant phase factor, so for our purposes, what's true of one is true of the other. We will therefore regard the reduced parametrization when showing that the solution is a Nash equilibrium in the next section.

If we further reduce the parametrization by letting all phase parameters $\alpha_1,\alpha_2,\alpha_3,\beta_1,\beta_2=0$, we end up with an operator $O(\phi,\theta,\chi) \in$ SO(3), i.e. the elements of the special orthogonal group of dimension 3. These operators corresponds to rotations in $\mathbb{R}^3$. In quantum games with two choices, like quantum prisoners dilemma and in minority games, local \emph{orthogonal} operations merely achieves to replicate the results of classical mixed strategies and offers no improvement in the expected payoff, even with a maximally entangled initial state. In this case though, there exists an $O^{opt} \in$ SO(3), given in table 3 that outperforms the classical expected payoff by a small margin. Each player would in this case receive a payoff of  $E(\$)=\frac{40}{81}$, compared to the classical $E(\$)=\frac{4}{9}=\frac{36}{81}$. This result might open up the possibility of a new classification of quantum games, where there could exist a category of quantum games with classical strategies that are fundamentally different than classical games with classical strategies.

\begin{table}[!h]\center
\label{Uopttable}
\begin{tabular}{|c|c|c|c|c|c|c|c|c|c|}
\hline
\hline
 Parameter  & $\phi$ & $\theta$  & $\chi$ \tabularnewline

\hline
 Value  & $\frac{\pi }{6}$ & $\cos ^{-1}\left(\frac{1}{3}\right)$ & $\frac{\pi }{6}$\tabularnewline
\hline
\end{tabular}
\caption{$O^{opt}$ in the given parametrization.}
\end{table}

\subsection{Nash equilibrium}
To show that this solution is valid from a game-theoretical point of view, we need to show that $V_{opt}$ is a Nash equilibrium, i.e. that none of the players gains by unilaterally changing strategy from $V^{opt}$ to any other strategy $U(\phi,\theta,\chi,\alpha,\beta_1,\beta_2)$. Without loss of generality, we show for the expected payoff $E_A(\$)$ of Alice that the following inequality holds for any $V$:
\begin{equation}
E_{A}(\$)(V_{C}^{opt}\otimes V_{B}^{opt}\otimes V_{A}^{opt})\geq E_{A}(\$)(V_{C}^{opt}\otimes V_{B}^{opt}\otimes V_{A}).
\end{equation}
We show that this is the case by letting Alice act with a general unitary operator $V(\phi,\theta,\chi,\alpha,\beta_1,\beta_2) \in$ SU(3), while Bob and Charlie acts with $V^{opt}\; (U^{opt})$. Then we take the partial derivatives of $E_A(\$)$ with respect of each of the parameters while keeping the rest at optimal value. Vanishing partial derivatives together with a negative definite Hessian matrix at the values of $V^{opt}$ proves that $V^{opt}$ is Alice's dominant strategy and because of the symmetry of the protocol, thereby a Nash equilibrium.

\begin{multline}
\left.\frac{\partial E_{A}(\$)}{\partial\phi}\right|_{\phi=\phi^{'}}=\left.\frac{2}{9}\cos(2\phi)\right|_{\phi=\phi^{'}}=0,\qquad
\left.\frac{\partial^{2}E_{A}(\$)}{\partial\phi^{2}}\right|_{\phi=\phi^{'}}<0,
\end{multline}

\begin{multline}
 \left.\frac{\partial E_{A}(\$)}{\partial\theta}\right|_{\theta=\theta^{'}}=\left.\frac{1}{27} \left(-\sqrt{3} \sin (\theta )+3 \sqrt{2} \cos (2 \theta )+\right.\right.\\
 \left.\left.\left(3 \sin (\theta )+\sqrt{6}\right) \cos (\theta )\right)\right|_{\theta=\theta^{'}}=0,\qquad\left.\frac{\partial^{2}E_{A}(\$)}{\partial\theta^{2}}\right|_{\theta=\theta^{'}}<0,
\end{multline}

\begin{multline}
\left.\frac{\partial E_{A}(\$)}{\partial\chi}\right|_{\chi=\chi^{'}}=\left.\frac{2}{27}(\cos(\chi)-\sin(\chi))\left(\sin(\chi)+\cos(\chi)+\sqrt{2}\right)\right|_{\chi=\chi^{'}}=0,\\\left.\frac{\partial^{2}E_{A}(\$)}{\partial\chi^{2}}\right|_{\chi=\chi^{'}}<0,
\end{multline}

\begin{multline}
\left.\frac{\partial E_{A}(\$)}{\partial\alpha}\right|_{\alpha=\alpha^{'}}=\left.\frac{4 \cos (\text{$\alpha$})}{27}\right|_{\alpha=\alpha^{'}} =0,\qquad\left.\frac{\partial^{2}E_{A}(\$)}{\partial\alpha^{2}}\right|_{\alpha=\alpha^{'}}<0,
\end{multline}




\begin{multline}
\left.\frac{\partial E_{A}(\$)}{\partial\beta_{1}}\right|_{\beta_{1}=\beta_{1}^{'}}=\left.\frac{1}{54}\left(-3\sin(\text{\ensuremath{\beta_{1}}})+\sin(2\text{\ensuremath{\beta_{1}}})+\right.\right.\\
\left.\left.+3\sqrt{3}\cos(\text{\ensuremath{\beta_{1}}})+\sqrt{3}\cos(2\text{\ensuremath{\beta_{1}}})\right)\right|_{\beta_{1}=\beta_{1}^{'}}=0,\qquad\left.\frac{\partial^{2}E_{A}(\$)}{\partial\beta_{1}^{2}}\right|_{\beta_{1}=\beta_{1}^{'}}<0,
\end{multline}

\begin{multline}
\left.\frac{\partial E_{A}(\$)}{\partial\beta_{2}}\right|_{\beta_{2}=\beta_{2}^{'}}=\left.\frac{1}{54}\left((2\sin(\text{\ensuremath{\beta_{2}}})+3)(-\cos(\text{\ensuremath{\beta_{2}}}))-\right.\right.\\
\left.\left.-\sqrt{3}(3\sin(\text{\ensuremath{\beta_{2}}})+\cos(2\text{\ensuremath{\beta_{2}}}))\right)\right|_{\beta_{2}=\beta_{2}^{'}}=0,\qquad\left.\frac{\partial^{2}E_{A}(\$)}{\partial\beta_{2}^{2}}\right|_{\beta_{2}=\beta_{2}^{'}}<0.
\end{multline}

\vspace{6 mm}

By calculating the Hessian $H$ with
\begin{equation}
\left. H_{ij}=\frac{\partial^2}{\partial a_i \partial a_j} E_A(\$)\right|_{a_{i}=a_{i}^{opt},a_{j}=a_{j}^{opt}},
\end{equation}
 where $a_i,a_j \in \{\phi,\theta,\chi,\alpha,\beta_1,\beta_2\}$, and confirming that all eigenvalues are negative, we conclude that $V^{opt} \;(U^{opt})$ is indeed a Nash equilibrium.

\subsection{Adjusting entanglement and fidelity}
We have included a simple model of noise, to show the behavior of the expected payoff, when the initial state was adjusted towards a completely mixed state. This was done by controlling the fidelity $f$ of the initial state, by mixing it with an even distribution of all basis states in $\mathcal{H}=\mathbb{C}^{3}\otimes\mathbb{C}^{3}\otimes\mathbb{C}^{3}$. Clearly as $f\rightarrow0$, we should expect the entanglement as a resource in the initial state to vanish. This is of course the case and we have $E(\$)(U^{opt},f)=(2 (2 + f))/9$. For $f=0$ we simply end up with the classical result.

A way of directly adjusting the strength of entanglement in the initial state, while keeping the state pure is to start with
\begin{equation}
\mid\psi_{in}\rangle=\sin\vartheta\cos\varphi|000\rangle+\sin\vartheta\sin\varphi|111\rangle+\cos\vartheta|222\rangle,
\end{equation}
where $0\leq\vartheta\leq\pi$ and $0\leq\varphi\leq2\pi$. We retrieve the maximally entangled state (\ref{eq:GHZ}) for $\varphi=\frac{\pi}{4},\frac{3\pi}{4}$  and $\vartheta=\pm\cos^{-1}(1/\sqrt{3})$.  The expected payoff is given by:
\begin{equation}
E(\$)(U^{opt},\vartheta,\varphi)=\frac{1}{9}\left(\sin(\varphi)\sin(2\vartheta)+\cos(\varphi)\left(2\sin(\varphi)\sin^{2}(\vartheta)+\sin(2\vartheta)\right)+4\right),
\end{equation}
which shows that any deviation from maximal entanglement reduces the expected payoff towards the classical $E^{c}(\$)$, graphically shown in figure 1.

A point to note here is that the maximum expected payoff strongly depends on the choice of initial state $|\psi_{in}\rangle$, and that there can exist more or less suitable initial states depending on the task. We chose the GHZ-state for this protocol because it is an unbiased maximally entangled state, which lets the classical game be present and accessible trough restrictions on $S$. Would our preferences been different and we had chosen for example the antisymmetric Aharonov state instead:
\begin{equation}
|\mathcal{A_{-}}\rangle=\frac{1}{\sqrt{6}}\sum_{x_3,x_2,x_1=0}^{2}\epsilon_{x_{3}x_{2}x_{1}}|x_{3}x_{2}x_{1}\rangle,
\end{equation}
where $\epsilon_{x_{3}x_{2}x_{1}}$ is the completely antisymmetric tensor, then the expected payoff would have been $E(\$)=1$, just by letting the players apply the identity operator. This state would guarantee that everybody ends up with a unique choice every time. But that wouldn't be of any interest from a game theoretical perspective since outcomes would have resembled a classical game with unrestricted communication. However, due to the the invariance of $|\mathcal{A_{-}}\rangle$ under local unitary transformations of the form $U\otimes U\otimes U$, superpositions of $|\mathcal{A_{-}}\rangle$ and $|000\rangle$ under some restricted set of operators resembling the set of mixed classical strategies, could model a classical game under different amounts of communication.
\begin{figure}
\center
\includegraphics[scale=0.55]{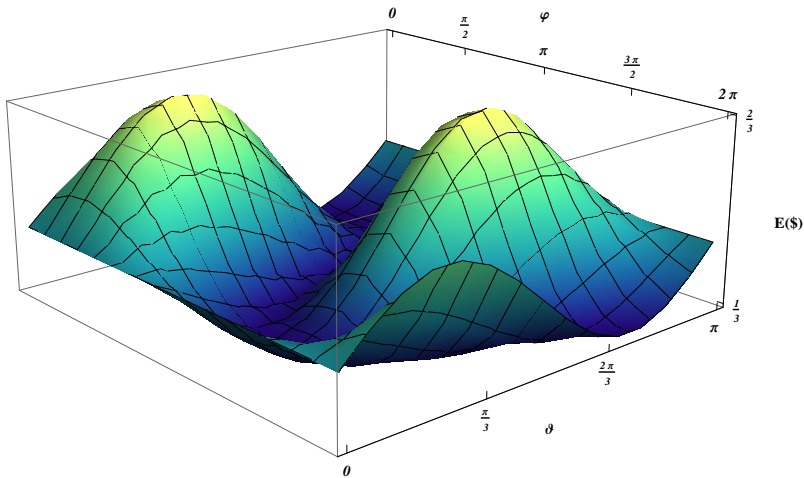}
\caption{Expected payoff $E(\$)(U^{opt},\vartheta,\varphi)$ as a function of $\vartheta$ and $\varphi$ at the Nash equilibrium strategy.}
\end{figure}

\section{Conclusions}
We have created the first quantum model for a three player, three restaurant Kolkata restaurant problem. We have shown that when the players share an initial tripartite entangled state, there exists a local unitary operation for which the players can increase their expected payoff $E(\$)$ by 50\% compared with classical randomization. This solution is a Nash equilibrium and therefore a natural attractor in the space of available strategies.  The achievement of this performance is highly dependent on the strength of entanglement and the fidelity of the initial state.

\begin{flushleft}
\textbf{Acknowledgments:}
We wish to thank Ole Andersson for valuable inputs and fruitful discussions.
This  study was supported  by the Swedish Research Council (VR).
\end{flushleft}

\end{document}